# POWER AWARE PHYSICAL MODEL FOR 3D IC'S

Yasmeen Hasan

Dept Of ECE, Integral University, Lucknow,India

Email: yasmeen.hasan9@gmail.com

**Abstract**- In this work we have proposed a geometric model that is employed to devise a scheme for identifying the hotspots and zones in a chip. These spots or zone need to be guarded thermally to ensure performance and reliability of the chip. The model namely continuous unit sphere model has been presented taking into account that the 3D region of the chip is uniform, thereby reflecting on the possible locations of heat sources and the target observation points. The experimental results for the – continuous domain establish that a region which does not contain any heat sources may become hotter than the regions containing the thermal sources. Thus a hotspot may appear away from the active sources, and placing heat sinks on the active thermal sources alone may not suffice to tackle thermal imbalance. Power management techniques aid in obtaining a uniform power profile throughout the chip, but we propose an algorithm using minimum bipartite matching where we try to move the sources minimally (with minimum perturbation in the chip floor plan) near cooler points (blocks) to obtain a uniform power profile due to diffusion of heat from hotter point to cooler ones.

## 1: Introduction

In the recent years the power density of integrated circuits(IC's) has doubled every three years. Because energy consumed by the chip is converted into heat, this continuing exponential rise in power density creates vast difficulties in cooling costs [1][2]. The most critical challenge of 3D IC design is heat dissipation, which has already been realized and studied even for 2D IC designs. There have been several existing works on 3D power and temperature aware physical designs and management. In order to device a scheme for identifying the hotspots and zones in a chip S.Majumder and S.S.Kolay[3] proposed two different geometric models , namely continuous and discrete to take into account whether the 2D plane of the chip floor is gridless or a uniform grid, thereby reflecting on the possible locations of the heat sources and target observation points. These spots or zones need to be guarded thermally to ensure performance and reliability. Some researchers have addressed the problem of identification of hotspots in VLSI chips [4] whereas others have proposed an alternative placement scheme to cool down a hot chip[5].Kang has highlighted the need for new thermal designs in order to cope up with the challenging scenarios ensuing from the high-scale integration[6]. Miranda et al. had successfully used the finite volume method approach to investigate maximum temperature rising on a CPU motherboard [7]. Jing Li and Hiroshi Miyashita (2006) [8] proposed a finite difference thermal Model. Where it was assumed that every heat source that overlaps the effective area Aeff of a grid point serves as a power source feeding into the grid and the corresponding power value of the grid is calculated based on the ratio of the source area within Aeff to the total area of the source.

In this paper we have proposed a geometric model which is employed to devise a scheme for identifying the hotspots in a three dimensional integrated circuit(IC).

We propose a model here which may facilitate in identifying the hot spots/zones in a VLSI chip. In the continuous domain we have used the concept of a unit sphere model to calculate the local thermal effect at a point due to the heat being dissipated from several point heat sources distributed over the chip plane. We establish that a point on a chip can become very hot due to the conduction effects of other heat

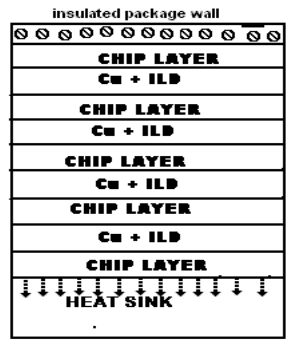

**Fig1:** 3D IC technology

sources, although it may not have a heat source in its immediate vicinity. In this model, the heat loss due to radiation has been ignored. If it is to be considered, an appropriate heat loss has to be incorporated functions

### 2.1: Time Invariant Heat Sources

The study is made with the assumption that there are constantly active (i.e. always on) heat generating sources placed randomly throughout the chip. For continuous thermal sources; we also assume that the heat from the sources is being propagated through the 3D surface of the chip without being dissipated in the ambience. The objective is to identify the zones in the chip, which have heat content greater than a certain threshold.

### 2.2: Continuous Spatial Domain

The position of a heat source may be any point on the chip which is assumed to be a 3D integrated circuit (IC). In the unit sphere model, the contribution of a point heat source S at any target point T is expressed as the amount of heat from S received within the unit sphere centered at the point T. This unit is the same as that of the distance between S and T, and may be related to the minimum dimension of the chip. The cumulative heat received at the point T is evaluated as the linear superposition of the amounts received at T from all heat – generating sources on the chip. As illustrated with Fig. 2, let a heat source at a point S generate an amount Q, henceforth denoted as the strength of the source S. Let the target point T be at a Euclidian distance d from S. Let $C_T$ and $C_s$ intersect at the two points A and B.

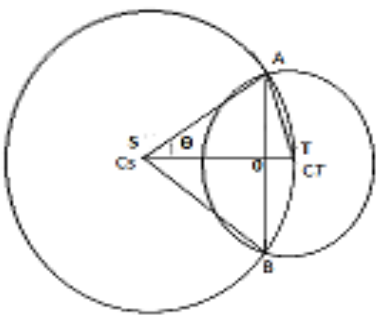

Fig 2: Unit Sphere Model of Heat Received at a Point T

Then the area cut out on the surface of the sphere $C_S$ is equal to the product of solid angle with its vertex at the center of the sphere $C_s$ and the square of the sphere's radius.

$$A = \tau \times d^2 \quad \quad \ldots (1)$$

Where $\tau$ *is the solid angle* formed by the conical surface of the spherical sector and d is the radius of the source sphere [10].

A complete sphere forms a solid angle of $4\pi$. (If the solid angle is not formed by the entire sphere, but only by a conical surface of a spherical sector, the angle in this case is equal to the ratio of the sector's spherical surface to the square of the sphere's radius [11][12].)

By denoting the plane angle at the vertex of the spherical sector as θ, it is possible to express its height h as

$$h = d - d\cos\theta = d(1 - \cos\theta) = 2d\sin^2\frac{\theta}{2} \qquad \ldots (2)$$

where r is the radius of the source sphere.

Therefore the spherical area of the sector can be represented as

$$A = 2\pi d^2(1 - \cos\theta) = 4\pi d^2 \sin^2\frac{\theta}{2} \qquad \ldots (3)$$

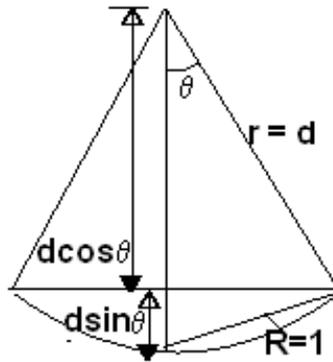

Fig 3: Section of a cone and a spherical cap inside a sphere

By denoting the solid angle which subtends the spherical surface of the sector as we obtain

$$\tau = 4\pi\sin^2\frac{\theta}{2} = 2\pi(1 - \cos\theta) \qquad \ldots (4)$$

Thus the contribution of heat from S at T is

$$Q \times \frac{2\pi d^2(1-\cos\theta)}{4\pi d^2} \qquad \ldots (5)$$

Where   is the surface area of the sphere S.

Consider $OC_TB$ in the figure 10

$(C_TB)^2 = (OB)^2 + (OC_T)^2$

$$1 = 2d^2(1-\cos\theta) \qquad \ldots (6)$$

Putting eqn (6) in eqn (5) we get

The contribution of heat from S to T is

$$= Q \times \frac{1}{4d^2} \qquad \ldots (7)$$

Our concerns are the hottest points on the chip. Intuitively, the source points definitely belong to the above class. But the more pertinent question is whether these are the only points that need to be considered. The question may be re-phrased as follows: does there exist any non-source point on the floor with heat content greater than that of any of the source points?

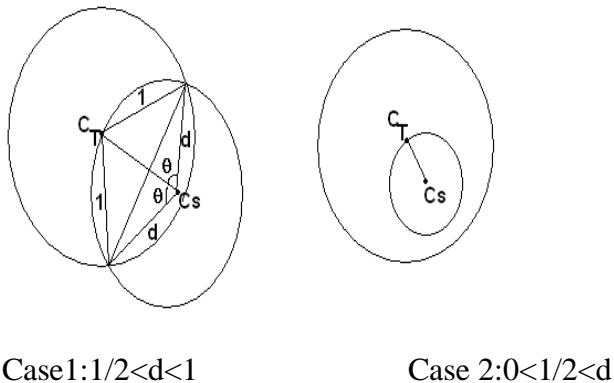

Case1:1/2<d<1    Case 2:0<1/2<d

Fig 4:  Special cases of the unit sphere model

The observations reported, answer in the affirmative. Before we proceed further, we point out two special cases of the unit sphere model fig 4 based on the distance d between S and T:

0.5<d<1     and      (2) 0<d<0.5

In the boundary case when S lies on $C_T$ is equal to   , as SAT becomes an equilateral triangle

$$Q = \frac{Q_o}{2}(1-\cos\theta)$$

$$= \frac{Q_o}{2}\left(1 - \frac{1}{2}\right) \quad = \frac{Q_o}{4} \quad \ldots (8)$$

Hence in case (1) the angle 2θ as defined earlier will be greater than, and consequently more than   of the heat emanating from S reaches the unit sphere centered at T. In case (2) T is nearer to S and hence the sphere with radius 'd' around S will now lie entirely within the Unit sphere at T. Hence the unit sphere $C_T$ receives the entire heat   of S in this case.

## 2.3 : Effective thermal conductivity

### Conductivity Of Different Layers In 3D IC Is Different

As we know that in a multilayered 3D IC the thermal conductivity of the dielectric layers inserted between device layers for insulation is very low compared to silicon and metal. For instance, the thermal conductivity for epoxy, kepoxy is 0.05W/mK, while ksilicon is 150 W/mK and kcopper is 285 W/mK.

Since the thermal conductivity of each layer in a multilayered 3D IC is different from each other, hence the propogation of heat through each layer is different.

In our model we assume that there is uniform propagation of heat in all the directions, but this is not the case in a 3D IC and hence we might end up considering a given target point much more hotter than it actually is. Thus we can use the concept of **"Effective thermal conductivity".**

For a multilayered structure we can define " effective thermal conductivity" [16] in the form as given below

$$wK_{eff} = \sum_i wi \times Ki \quad \ldots (9)$$

in W/K. Where the summation are applied over all the layers. Thermal conductivity can be considered effective also in the sense, that the patterned layers contribute to heat conductivity only partially, depending on their patterns ($w_i$= thickness of the layers).

After calculating the power P using the continous model formula at a point we can calculate the temperature at that point using the formula given as

$$T = \frac{P}{wK_{eff}} \quad \ldots (10)$$

The effective thermal conductivity can be calculated as follows*:*

If the thermal conductivity of silicon--resin--oxide layers are $K_{si}$= 0.72W/cmK -- K res=0.0063W/cm -- $K_{ox}$= 0.017W/cmK respectively and the thickness of the respective layers is $W_{si}$-- $W_{res}$ -- $W_{ox}$ in mm, then the value of effective thermal conductivity is given as

$$wKeff = W_{si}*Ksi + W_{res}*K_{res} + W_{ox}*K_{ox} \quad \ldots(11)$$

## 3: EXPERIMENTAL RESULTS

# RESULTS FOR THE CONTINUOUS DOMAIN

We performed more experiments in the continuous domain model implemented in C to simulate the effect of active sources placed at random points on the 3D floor. Keeping the dimensions of the 3D structure the same we varied the number of sources from 5 to 50.We have studied five trail runs, keeping the number and range of the power strength of the active sources fixed, just allowing the position of the sources to vary.

We observed that the results did not depend much on the randomness of the position of the active sources. As in the previous experiment we have observed that in the continuous domain the relatively hot points lie near the active source.

During simulation we considered more of those points for evaluation of the cumulative power. We actually considered a fine grid around each source point and evaluated the cumulative power at each of those points along with the source points. Also across the whole floor we considered a relatively coarse grid and evaluated the power at all the grid points of this coarse grid.

In the result we also found the coordinates and cumulative power of the target points. Here we observed that there were some target points whose cumulative power exceeded the cumulative power of the active source nearest to it. The result of our experiments confirm that in 3D IC 's as the number of power sources increases there is an increase in the power density.This means that in 3D IC's there is a need of efficient power management techniques.

## .4: CONCLUSION

In this work we have proposed a model in the continuous domain to model the thermal behavior in a 3D VLSI chip. The hotspots were usually concentrated near the active source points, but some points away from the source were found to be much hotter than the sources itself. The randomness of the source did not affect the result much. One important aspect we have observed in all the models is that there are zones in the chip which become much hotter even without containing a heat source. We conclude that it may not be enough to guard only the active regions to make the chip thermally stronger. This also requires the need for more efficient power and thermal management techniques[13][14][15].

We also observed that the numbers of hotspots reported in our model were very high compared to those found in the 2D case[3], which shows that the problem of increased power density with the increase in number of sources becomes more grave in the 3D case. Hence we discuss some power management methods to solve the increased power density problem and we also try to propose an algorithm using minimum bipartite matching where we try to move the sources minimally (with minimum perturbation in the chip floorplan) near cooler points (blocks) to obtain a uniform power profile due to diffusion of heat from hotter point to cooler ones, thus reduce the maximum temperature.

### 4.1: Power Management Techniques

The result of our experiments confirm that in 3D IC 's as the number of power sources increases there is an increase in the power density of the chip .This means that in 3D IC's there is a need of efficient power management techniques. The availability of power management features impacts partitioning and floorplanning of the system to reach power goals. The parameters on which power depends directly imply the types of features that can be used for power management. Switching off and scaling clocks to reduce dynamic power, switching off and scaling supplies to reduce dynamic power and leakage to a small extent, and switching of supplies and scaling threshold voltage to reduce leakage power are among the key techniques to manage power.

To save power, clock gating refers to adding logic to a circuit to prune the clock tree, thus disabling portions of the circuitry where flip-flops do not change state. A chip can have multiple clock domains each operating at a clock frequency as required to meet application schedule. Clock synchronization will be needed on signals that cross these clock domain boundaries. Power gating (PG) is a technique for eliminating leakage power consumption of unused blocks in certain modes of chip operation.

| NO OF SOURCES | THRESHOLD VALUE | TOTAL PROBES POINTS | PROBES POINTS IN FM | PROBE POINTS IN CM | HOTSPOT IN FM | HOTSPOT IN CM | %HOTSPOT IN FM | %HOTSPOT IN CM |
|---|---|---|---|---|---|---|---|---|
| 5 | 1.24876 | 2029160 | 29160 | 2000000 | 9198 | 2562 | 0.45% | 0.19% |
| 10 | 1.24338 | 2058320 | 58320 | 2000000 | 16798 | 8376 | 0.81% | 0.41% |
| 20 | 1.26821 | 2116640 | 116640 | 2000000 | 42238 | 12048 | 1.72% | 0.68% |
| 40 | 1.26441 | 2233280 | 233280 | 2000000 | 93972 | 19030 | 4.21% | 0.82% |
| 50 | 1.20101 | 2291600 | 291600 | 2000000 | 118262 | 25969 | 5.16% | 1.13% |

TABLE 1: RESULTS FOR THE CONTINUOUS DOMAIN

In multiple supply voltages method different blocks of a chip can operate at different voltages based on performance requirements; voltage supplies can be always-on or turned off. Each partition is associated with a fixed operating voltage. There may be an application specific accelerator on the chip that needs to operate at much higher frequency whereas the rest of the chip can operate at a lower frequency. Most of the chip can make use of a lower voltage compared to the accelerator part and reduce dynamic power consumption significantly. Even the leakage in active mode will be reduced due to the use of a lower supply voltage on the majority of the chip. The maximum frequency at which the design can run safely decreases with decreasing voltage. Thus, the system can reduce processor energy consumption by reducing design voltage, but this necessitates running the system at a slower speed. This implies that voltage can also be scaled to meet performance needs.

From a power management concept standpoint, AVS(Adabtive voltage scalling) is similar to DVS(dynamic voltage scalling). However, unlike DVS, which uses table lookup, AVS is a closed loop system and the power controller interfaces with a monitor in the scaled block to determine frequency needs and then directs the system to provide appropriate voltage.

Active back bias (ABB) voltage, applied to wells of N-MOS and P-MOS transistors, is used to set the threshold voltages and leakage currents precisely in order to improve speed and at the same time control device sub-threshold leakage. The active back bias applies a voltage to the well of devices and this voltage can be generated by a PMIC. If the leakage increases with age, temperature or other conditions, changes in bias supply can be used to compensate. Compensation is required for process variations also; ABB is likely to become a necessity at 45 nm and below.

Thermal vias can make possible the heat flow between the top area of the die and the underside of the lower copper plane for a small region centered about one of the vias. Since the vias have copper, they provide a path of least resistance and heat is transferred through the vias in a proportion much greater than the area of the vias. Vias can also provide both electrical and thermal path.

Synthetic jets offer an attractive solution for highly efficient localized cooling of integrated circuits. These jets are formed by time-periodic, alternate suction and ejection of fluid through an orifice bounding a small cavity, by the time periodic motion of a diaphragm that is built into one of the walls of the cavity. Unlike conventional jets, synthetic jets are "zero net mass flux" in nature and produce fluid flow with finite momentum with no mass addition to the system and without the need for complex plumbing. Because of their ability to direct airflow along heated surfaces in confined environments and induce small-scale mixing, these jets are ideally suited for cooling applications at the package and heat sink levels

**4.2: The Minimum Weight Bipartite Matching Problem**

The minimum weight bipartite matching problem occupies a central position in combinational optimization, and a variety of applications to transshipment problems.

The problem is formally defined as follows:

Obtain a minimum weight perfect matching in an edge-weighted bipartite graph.

1) Let T denote the threshold value. The threshold value is the minimum of the cumulative power of all source points.

2) Let A = [ ]  denote a matrix giving the power density at each of the grid points

3) We calculate

   $T - A = S = [\ ]$ = Excess at each grid point

   Where S denotes a matrix giving the excess at each grid point

   (Some grid points will have positive excess, some negative excess and some zero excess)

4) Identify the sources with their cumulative power. Let the sources be placed at $(x_s, y_s)$.

   Perform minimum weight bipartite matching

   Given: A bipartite graph, G= (V, E) where $V = V_1 \cup V_2$

   $V_1$ = Set containing the sources;

$V_2$ = Set containing all empty grid (target) positions having Excess $\geq$ [max (excess of source points)]

For each $e \in E$ and e= $(v_i, v_j)$, where $v_i \in V_1$ and $v_j \in V_2$

For an edge e ($v_i$, $v_j$) ∈ E from each vertex $v_i$ ∈ $V_1$ to all vertices $v_j$ ∈ $V_2$

Wt (e) = weight is minimum Manhattan distance from the original position of each source $v_i$ ∈ $V_2$ to the grid corresponding to $v_j$ ∈ $V_2$

A min weighted bipartite matching is performed to obtain assignment of each source $v_i \in V_1$ to a target new location $v_j \in V_2$.

The running time of the minimum weighted bipartite matching algorithm[25] is $O(m\sqrt{n}) = O(n^{2.5})$

However a detailed study of the algorithm remains an area of future work.

In this work we have considered a uniform propagation of heat in all the directions. But due to the different thermal conductivities of the different layers, if we take into consideration this non uniform propagation of heat in different layers, we can obtain a more accurate power distribution profile in the model. We have done some work in this regard but a detailed study remains an area for future work.

## Author

Yasmeen Hasan has received the MTech degree from Integral University, Lucknow, in 2009 in Electronic circuits & systems(VLSI). She is currently an Asstt Professor of Electronics and Communication at Integral University,Lucknow,India. Her research interest includes various aspects of Power and Temperature aware VLSI.

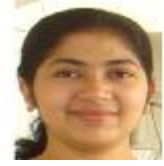